\documentclass[letterpaper, 10 pt, conference]{ieeeconf}  

\IEEEoverridecommandlockouts                              

\usepackage{graphicx}
\usepackage{subcaption}
\usepackage{multirow}
\usepackage{amssymb}
\usepackage{amsmath}
\title{\LARGE \bf
PPTP: Performance-Guided Physiological Signal-Based Trust Prediction in Human-Robot Collaboration 
}

\author{
Hao Guo, Wei Fan, Shaohui Liu, ~\IEEEmembership{Member,~IEEE}, Feng Jiang, ~\IEEEmembership{Senior Member,~IEEE}, Chunzhi Yi, ~\IEEEmembership{Member,~IEEE}
}

\begin{document}

\maketitle
\thispagestyle{empty}
\pagestyle{empty}

\begin{abstract}

Trust prediction is a key issue in human-robot collaboration, especially in construction scenarios where maintaining appropriate trust calibration is critical for safety and efficiency. This paper introduces the Performance-guided Physiological signal-based Trust Prediction (PPTP), a novel framework designed to improve trust assessment. We designed a human-robot construction scenario with three difficulty levels to induce different trust states. Our approach integrates synchronized multimodal physiological signals (ECG, GSR, and EMG) with collaboration performance evaluation to predict human trust levels. Individual physiological signals are processed using collaboration performance information as guiding cues, leveraging the standardized nature of collaboration performance to compensate for individual variations in physiological responses. Extensive experiments demonstrate the efficacy of our cross-modality fusion method in significantly improving trust classification performance. Our model achieves over 81\% accuracy in three-level trust classification, outperforming the best baseline method by 6.7\%, and notably reaches 74.3\% accuracy in high-resolution seven-level classification, which is a first in trust prediction research. Ablation experiments further validate the superiority of physiological signal processing guided by collaboration performance assessment.

\end{abstract}

\section{INTRODUCTION}

Recent advancements in robotics have expanded the possibilities for human-robot collaboration (HRC) in shared workspaces, particularly in construction \cite{liu2021brainwave,zhang2023human}, assembly \cite{zhang2022human}, and healthcare \cite{ait2024advancing}. Establishing appropriate levels of human trust in robots is essential for successful collaboration in these contexts \cite{azevedo2021unified}. When properly calibrated, trust enables focused task execution and improves collaborative efficiency, while inappropriate trust levels can impair performance \cite{chen2018planning,chen2020trust}. Consequently, accurately predicting trust dynamics under varying conditions is critical for effective HRC applications in construction. 

Despite trust's recognized importance, two primary challenges persist in its prediction and application. First, existing research has inadequately addressed construction-specific scenarios where humans and robots function as equal teammates \cite{emaminejad2024assessing}. Previous studies have typically focused on either human-dominated operations with robot assistance (e.g., operational guidance \cite{gupta2020measuring}, tool delivery \cite{chauhan2024predicting}) or robot-dominated operations with human supervision (e.g., autonomous driving \cite{lazanyi2017dispositional}). Chauhan et al. \cite{chauhan2024predicting} investigated construction scenarios but positioned robots merely as material providers rather than collaborative partners. As a result, existing trust models insufficiently capture trust dynamics in environments where humans and robots function as equivalent teammates.

Second, conventional questionnaire-based assessment methods provide only discrete measurements of trust before and after task completion \cite{khavas2021review}. The intricate interleaving of human and robotic execution steps in construction tasks, combined with variable operating conditions, necessitates continuous monitoring of trust dynamics. Physiological signals offer a promising alternative to overcome the limitations of questionnaire assessments \cite{golgouneh2020fabrication}. This is due to their continuous nature and demonstrated correlation with human cognitive states. 

Current physiological signal-based methods predominantly focus on isolated trust factors, neglecting the multifaceted nature of trust in complex collaborative environments. While numerous studies have validated the use of electroencephalography (EEG), electrocardiography (ECG), and galvanic skin response (GSR) signals for trust assessment, they face two critical limitations. First, substantial individual variations in physiological responses create inconsistencies \cite{chauhan2024predicting,akash2018classification,liu2021brainwave}. There are no standardized metrics to calibrate the relationship between these signals and trust levels. Second, many current approaches, such as EEG-based methods, heavily rely on laboratory equipment impractical for deployment in actual construction settings. Studies are often conducted in virtual reality environments rather than authentic construction scenarios. These limitations significantly hinder the development of robust, field-applicable trust prediction models for HRC in construction scenarios. 

To address these limitations, we designed a human-robot collaborative construction scenario with three difficulty levels, where participants alternated steps with a robot to complete block-based constructions. We collected synchronized data from multiple sources: construction images, ECG, GSR, and electromyography (EMG) signals. These data types were selected for their minimal deployment requirements. We also collected questionnaire responses correponding each task step. Based on this methodology, we propose a Performance-guided Physiological signal-based Trust Prediction (PPTP) framework. This framework integrates both individualized physiological measurements and standardized collaboration performance (CP) assessments. It processes physiological signals while using CP assessment as guiding information. Our approach achieves 81.1\% accuracy in three-level trust classification and 74.3\% accuracy in seven-level classification. 

The main contributions of this paper are:

\begin{itemize}
\item We investigate HRC trust in construction scenarios where humans and robots function as equal teammates, departing from previous paradigms where one agent dominated the interaction.
\item We propose a novel fusion mechanism integrating physiological signals with CP assessment, achieving complementarity between personalized physiological data and consistent CP metrics.
\item We develop a multimodal fusion framework that utilizes CP assessment to guide physiological signal feature extraction for trust prediction.
\item We achieve high-resolution seven-level trust classification, advancing beyond previous cognitive state prediction tasks limited to three-level, and demonstrate the effectiveness of incorporating collaboration performance assessment in trust prediction.
\end{itemize}

\section{RELATED WORK}

\subsection{Trust in HRC}

Trust is essential for robots to successfully work with their human counterparts. It is defined as the attitude that an agent will help achieve an individual's goals in a situation characterized by uncertainty and vulnerability \cite{law2021trust}. Madhavan et al. \cite{madhavan2007similarities} emphasize the critical role of trust in HRC, where trust determines the delicate balance of human self-reliance and over-reliance on robots. As a dynamic construct in HRC, trust is challenging to gain, maintain, and calibrate. The factors influencing human trust typically fall into three categories: task factors, human factors, and robot factors \cite{khavas2021review}, while its development occurs at dispositional, situational, and learned levels \cite{hoff2015trust}. Trust is usually measured directly by subjective scales as a standard measure or by behavioral evaluation \cite{gebru2022review}. 

Trust mechanisms vary across different scenarios, leading to seemingly contradictory findings. While Lazanyi et al. \cite{lazanyi2017dispositional} found participants willingly relinquished control of autonomous vehicles in simple scenarios, Gupta et al. \cite{gupta2020measuring} observed increased trust in robots during complex tasks. In construction settings specifically, key trust factors include task difficulty, risk, cognitive load, task duration, time of error, and task performance \cite{gebru2022review}. Building on these insights, our study employs block-based construction scenarios with varying difficulty levels to examine trust variations. We integrate physiological measurements, collaboration performance assessments, and subjective scales to develop a comprehensive trust prediction model.

\subsection{Trust Assessment Methods}

Trust assessment methods generally fall into subjective and objective categories. Subjective scales (e.g., TOAST \cite{wojton2020initial}, MDMT \cite{malle2021multidimensional}) serve as standard trust measurements but are limited to discrete evaluations before and after tasks, failing to capture dynamic trust changes during execution. Objective assessments include two approaches. The first evaluates interaction behaviors such as task intervention and opinion adoption. However, the fatal drawbacks of this type of approach are the need to operationalize behaviors \cite{lee1994trust,azevedo2021real} and differentiate between behaviors that indicate trust, and meanwhile that the attitudes, behaviors, and perceptions of people in a team are very different, as demonstrated by the studies of J. A. Cannon-Bowers and C. Bowers \cite{cannon2011team}, which leads to inherent differences in participants' behaviors matched to trust.

The second objective approach focuses on collaboration performance. Models like OPTIMo \cite{xu2015optimo} use dynamic bayesian networks to reduce uncertainty in operator trust estimation. Subsequent Bayesian models \cite{guo2021modeling,fooladi2021bayesian} have also applied inference over a history of robot performances. However, these models primarily rely on binary robot performance metrics (success/failure), neglecting other influential factors, also divorced from the influence of the human factor in HRC.

\subsection{Physiological signal-based Methods in Trust Studies}

Recent advances in machine learning have enhanced trust prediction using physiological signals. Chauhan et al. \cite{chauhan2024predicting} demonstrated this by manipulating interaction levels in virtual construction tasks and predicting trust using electrodermal activity (EDA), heart rate variability (HRV), and skin temperature with XGBoost and random forest algorithms. Similarly, Akash et al. \cite{akash2018classification} employed quadratic discriminant classifiers with EEG and GSR for real-time trust status discrimination.

For broader cognitive state assessment, researchers have explored various signal combinations. Liu et al. \cite{liu2023cognitive} developed the FS-MCCD algorithm for cognitive load prediction using EEG, EDA, ECG, EOG, and eye tracking in Open MATB tasks. Rajavenkatanarayanan et al. \cite{rajavenkatanarayanan2020towards} applied SVM to ECG and EDA data for cognitive load estimation during collaborative assembly. These studies establish the significance of physiological signals, particularly ECG and GSR, in cognitive state prediction. However, they share a critical limitation: reliance on individualized physiological responses without standardized performance metrics for reference, resulting in models performance lacking robustness.

Our approach addresses this limitation by incorporating ECG and GSR signals supplemented with EMG measurements. More importantly, we introduce a visual-based CP assessment function that provides a standardized assessment benchmark. This CP information serves as the guidance information in prediction process, effectively balancing human factors and task factors to enhance prediction accuracy.

\section{METHOD}

\begin{figure*}[t]
\centering
\includegraphics[width=\textwidth]{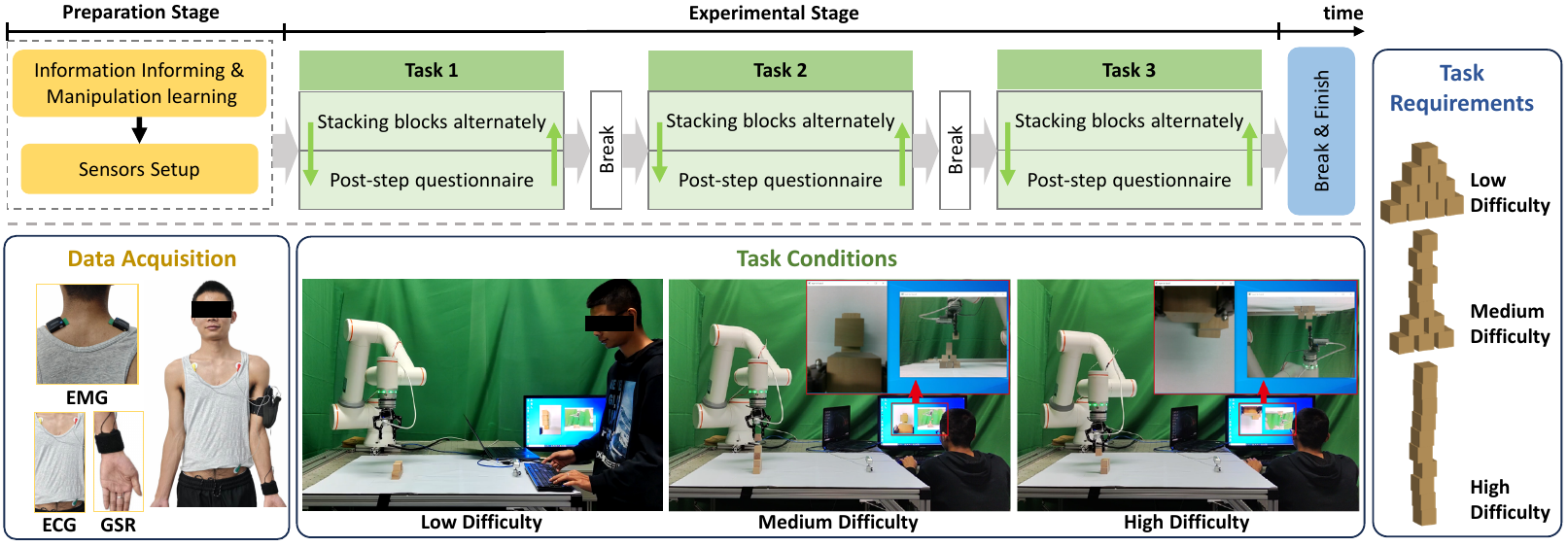}
\caption{Experimental Setup: Procedure Flow, Data Acquasition, Task Conditions, and Task Requirements Across Three Difficulty Levels.}
\label{fig1}
\end{figure*}

\subsection{Participants and Experiment Setup}

A total of 30 participants aged 18 to 32 were recruited for this study. The physiological data collection was approved by the Chinese Ethics Committee of Registering Clinical Trials. All participants signed consent forms before participating and were informed they could withdraw at any time. Our study employs the Fairino-FR16 robot as the collaborative teammate. Two cameras provided visual feedback: one in an eye-in-hand configuration for local views and another in an eye-to-hand position offering a global environmental perspective. During tasks, participants either directly observe the actual stacking environment or rely on camera feedback through a screen to assess the construction progress. 

\subsection{Experiment Design}
Participants were required to collaborate with a robot to build constructions alternately, using ten blocks per task. Each step involves stacking a single block using a robotic arm controlled either by participants via keyboard or by an autonomous manipulation system. Block detection and position evaluation for the autonomous system are performed by YOLO-v5 \cite{jiang2022review}, specifically trained for our study with an IoU accuracy exceeding 0.85. To preserve natural interaction, participants were not informed about the robot's performance capabilities. 

The experiment comprised three tasks with increasing complexity levels: low difficulty (LD), medium difficulty (MD), and high difficulty (HD). Fig. \ref{fig1} illustrates both the task conditions (observation methods) and task requirements (target constructions) for all three difficulty levels. Each difficulty level corresponded to a unique set of target construction and observation method as shown in Fig. \ref{fig1}. In the LD task, participants directly observed the stacking scene without camera view restrictions, enabling straightforward block manipulation. For the MD task, participants stacked blocks while relying exclusively on both eye-in-hand and eye-to-hand camera feeds displayed on a screen. The HD task introduced greater difficulty by inverting video frames from both cameras, creating significant disparity between visual perception and intuitive manipulation. To mitigate potential monotony effects, the sequence of these tasks was randomized for each participant.

Each task continued until all blocks are successfully stacked or the construction collapses. Participants could not physically touch the blocks, ensuring all manipulations occur through the robot. All blocks are identical in appearance to eliminate potential biases from color or size variations. Each manipulation step is time-bound to emphasize the cognitive load associated with varying task complexities.

\subsection{Data Acquisition and Preprocessing}
We collected multimodal physiological signals including ECG, GSR, and EMG. Data from preparation stages, questionnaire completion, and rest periods were excluded from analysis. Valid information included data from each task's execution stage and one minute post-task. All sensors were triggered by synchronized triggers to ensure temporal alignment of signals. EMG signals were acquired using Delsys Trigno sensors (sampling frequency: 1260Hz). ECG was recorded using an ADS1292R module, and GSR with a Grove GSR module (both at 125Hz). Due to temporal scale differences between signals (EMG manifests as millisecond-level muscle discharge activities \cite{9395610}, while GSR/ECG reflect second-level autonomic nervous system responses \cite{akash2018classification,liu2023cognitive}), we established differential window settings. We set 3-second windows for ECG and GSR, 216ms windows for EMG, with uniform 108ms overlap. All time windows were end-aligned in the temporal domain.

Participants' trust levels were captured using the Muir questionnaire \cite{muir1994trust,muir1996trust}, which served as ground truth for our physiological signal-based trust prediction model. Cognitive load was documented using the NASA-TLX questionnaire \cite{hart2006nasa}. Both questionnaires employed seven-point Likert scales ranging from "very low" to "very high." We collected Muir questionnaire data at each point when completing placing a block, establishing discrete trust measurements as reference standards. For continuous modeling purposes, we assumed constant trust levels between consecutive recording points. The average score from each questionnaire was used as the ground truth trust level for corresponding physiological signals collected between recording points. This approach provided the labeled trust data necessary for training and evaluating our prediction model. NASA-TLX scores were collected before and after each task to document participants' cognitive load levels and validate the effectiveness of our difficulty level manipulation.

\begin{figure*}[t]
\centering
\includegraphics[width=\textwidth]{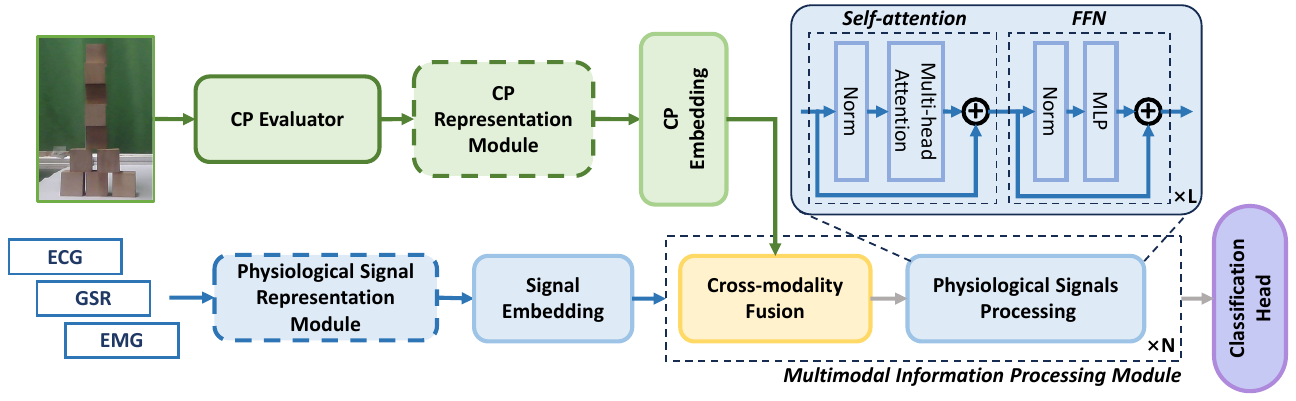}
\caption{Architecture of the Proposed PPTP Framework.}
\label{fig2}
\end{figure*}

\subsection{Model Structure}
The PPTP model proposed in this paper is shown in Fig. \ref{fig2} The model contains four main components: physiological signal representation module, collaboration performance evaluator, collaboration performance representation module and multimodal information processing module. The following subsections detail each component.

\subsubsection{Physiological Signals Representation Module}
The input physiological signals include ECG, GSR, and EMG of both trapezius muscles. After the 4-channel multimodal physiological signals are input to the representation module, EMG signals are individually processed by a FFN to match their length with the ECG and GSR signals. The uniformly-length physiological signals then pass through reversible instance normalization in their respective channels. Multimodal signals are broken into disjoint patches, and each patch is mapped to a multi-dimensional embedding using trainable linear projection. These embeddings serve as input for the subsequent process. 

\begin{figure}[htbp]
\centering
\includegraphics[width=2.5in]{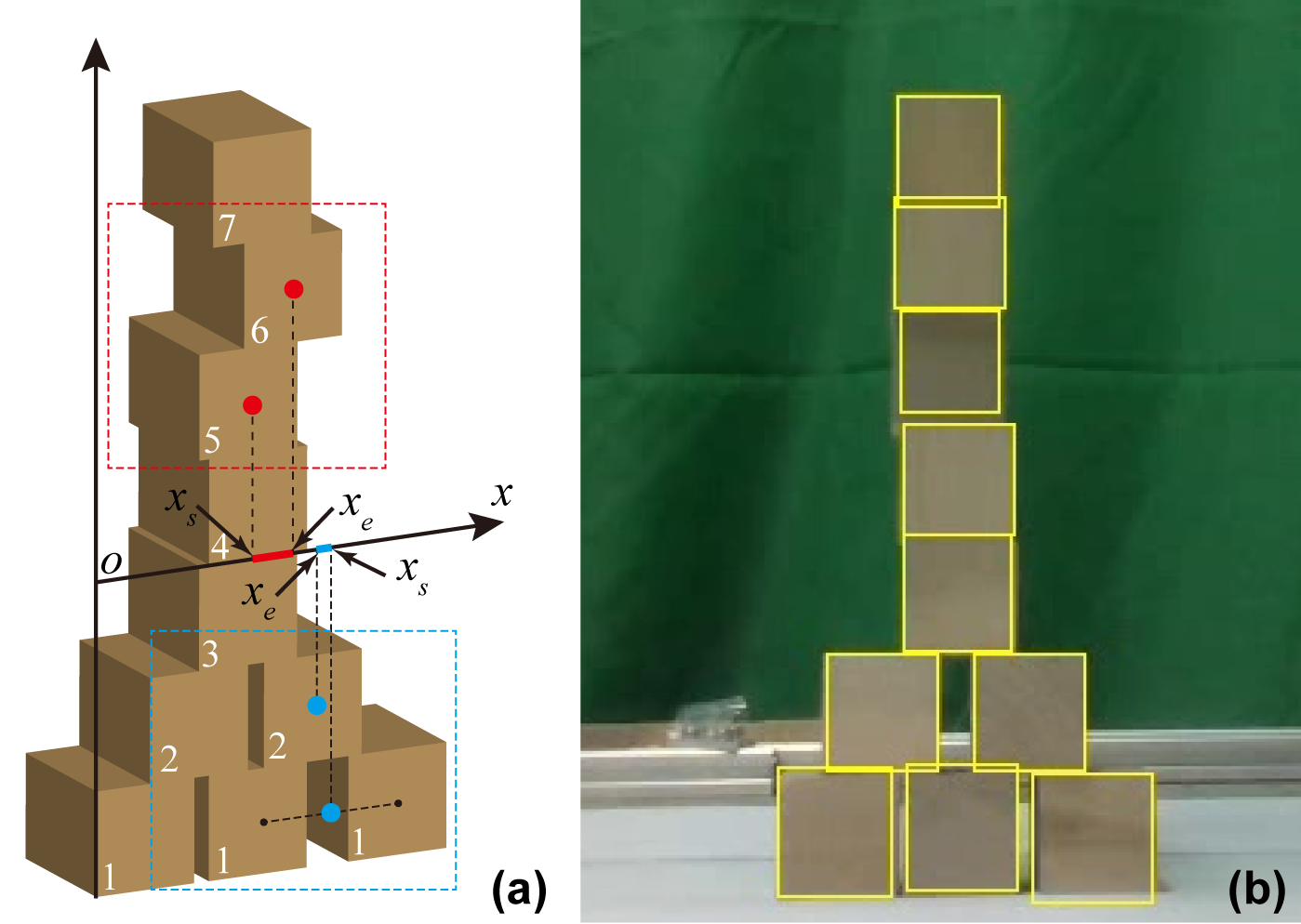}
\caption{Collaboration Performance Evaluation Process: Block Skewing Assessment and Failure Risk Analysis. The number marked at the corner of blocks represents the corresponding layer.}
\label{fig_3}
\end{figure}

\subsubsection{Collaborative Performance Evaluator and Representation}
Unlike previous binary success/failure evaluation methods, our performance evaluator first assesses the skewing of the current block, then combines it with historical performance to evaluate the entire construction. This approach quantifies performance levels more precisely than simple binary judgments. For a single block, we consider that higher-layer blocks have greater skewing impact on construction stability, as shown in (1):

\begin{equation}
S = \begin{cases}
|x_e - x_s|l, & l > 1\\
0, & l = 1
\end{cases}
\end{equation}

where $S$ represents the skewing of single block, the $x_e$ represents the horizontal coordinate of the evaluated block center, $x_s$ represents the horizontal coordinate of the support center under the evaluated block, l represents the evaluated block’s layer. Notably, the skewing for blocks at the base layer is zero, reflecting their stability. If the evaluated block is supported by two blocks (e.g. the part under the x-axis in Fig. \ref{fig_3}a, blue dots denote the evaluated block centers in the construction part), $x_s$ is calculated as the average horizontal coordinate of the centers of the two supporting blocks below. Conversely, if the evaluated block is supported by one block (e.g. the part above the x-axis in Fig. \ref{fig_3}a, red dots denote the evaluated block centers in the part stacked over the construction), $x_s$ corresponds to the center of the single block directly beneath it. Both $x_e$ and $x_s$ are detected and evaluated by YOLO-v5, as depicted in Fig. \ref{fig_3}b.

Considering that trust is heavily influenced by current CP and less by historical results, we utilize weighted moving average (WMA) \cite{perry2010weighted} to construct the failure risk vector $F \in R^{1 \times 10}$ that quantifies CP. We add position embeddings to $F$ to maintain temporal characteristics \cite{bello2019attention,parmar2018image}. This $F$ metric inversely represents construction quality by quantifying collapse risk at each building step. The evaluation function in (2) provides a continuous performance measure that extends beyond traditional binary success/failure judgments:

\begin{equation}
F_n = \begin{cases}
S_n + \gamma S_{n-1} + ... + \gamma ^ {n-1}S_1\\
\quad=\sum_{k=1}^n \gamma ^ {n-k} S_k, & n\leq 10\\
-1, & \text{unstacked}\\
-2, & \text{collapse}
\end{cases}
\end{equation}

where $F$ represents construction failure risk, $k$ represents the sequence of stacked blocks, $n$ represents the number of stacked blocks, and $\gamma \in$(0,1) represents the discount factor (set to 0.8). When a new block is stacked, the corresponding $F_n$ replaces the previous -1 value. When construction collapses, $F_n$ through $F_{10}$ are replaced with -2, while preceding values $F_1$ to $F_{n-1}$ remain unchanged.

\subsubsection{Multimodal Information Processing Module}
\begin{figure}[h]
\centering
\includegraphics[width=3in]{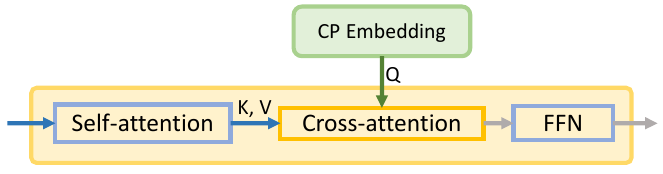}
\caption{Architecture of cross-modality fusion module.}
\label{fig_4}
\end{figure}

We developed a multimodal information processing module to combine CP and physiological signal features. The module contains two parts: physiological signals processing and cross-modality fusion. As depicted in Fig. \ref{fig2}, the physiological signals processing module builds upon modifications to the original Transformer \cite{vaswani2017attention}. Specifically, we remove additive bias from Layer Norm \cite{ba2016layer}, place it before residual skip connections \cite{he2016deep}, and use relative positional embedding \cite{shaw2018self}. Inspired by cross-modality feature alignment from vision-language models, we use CP as guiding information to direct attention to key segments in physiological signals. We insert cross-modality fusion into the deep processing of physiological signals, enabling more accurate feature selection. As shown in Fig. \ref{fig_4}, the key design difference in cross-modality fusion is the insertion of a cross-attention section between self-attention and FFN sections from the physiological signals processing module. This introduces guidance from CP into signal processing. In our implementation, we used L = 3 and N = (L+1)$\times$6 = 24 to ensure sufficiently deep processing of physiological signals and appropriate granularity of guidance features.

\section{EXPERIMENT}

\subsection{Statistical Scale Results}

We calculated trust scores based on the Muir questionnaire across three difficulty conditions. Interestingly, we obtained the lowest average trust scores under the LD condition, medium scores under the MD condition, and the highest scores under the HD condition. One-way ANOVA revealed significant differences (p$<$0.05) in average trust scores across the three difficulty conditions, aligning with our goal of inducing different levels of human trust in robots. We also conducted one-way ANOVA on cognitive load scores across the three difficulty conditions. Results showed that cognitive load scores followed the same trend as trust scores, with different difficulty conditions leading to significant cognitive load differences.

\subsection{Classification Performance of Individual Signals and Multimodal Integration}
TABLE \ref{tab:tab1} presents the classification results for 30 subjects using different signals. We first assessed each modal signal's independent classification performance for trust level prediction. Among the signals considered, GSR achieved the highest average accuracy of 77.3\%, while ECG demonstrated comparable performance of 77.1\%. These results align with previous findings that ECG and GSR strongly correlate with human intrinsic states. Notably, when evaluated separately, CP information outperformed EMG signals, confirming that our proposed CP metric contains effective information for differentiating dynamic trust levels.

\begin{table}[tb]
\caption{Comparison of Accuracy and F1 Score Across Single-modal and Full-modal Signals.}
\label{tab:tab1}
\begin{center}
\begin{tabular}{|l|cc|cc|}
\hline
\multirow{2}{*}{\textbf{Signals}} & \multicolumn{2}{c|}{\textbf{Accuracy}} & \multicolumn{2}{c|}{\textbf{F1 score}} \\
\cline{2-5}
& Mean & Std & Mean & Std \\
\hline
EMG & 70.4 & 8.8 & 58.9 & 13.1 \\
GSR & 77.3 & 2.9 & 71.2 & 7.6 \\
ECG & 77.1 & 2.9 & 71.1 & 5.7 \\
Collaboration Performance & 73.1 & 7.4 & 63.8 & 12.1 \\
All physiological signals & 78.5 & 3.4 & 73.1 & 7.4 \\
All physiological signals + CP& 81.1 & 3.6 & 75.9 & 7.1 \\
\hline
\end{tabular}
\end{center}    
\end{table}

When integrating all physiological signals, we compared classification performance with and without CP guidance. Results showed that integrating all physiological signals improved classification performance (78.5\% average accuracy) compared to individual signals. Furthermore, physiological signals guided by CP led to a significant improvement, achieving the highest average accuracy of 81.1\%.

\begin{figure}[bp]
\centering
\includegraphics[width=2in]{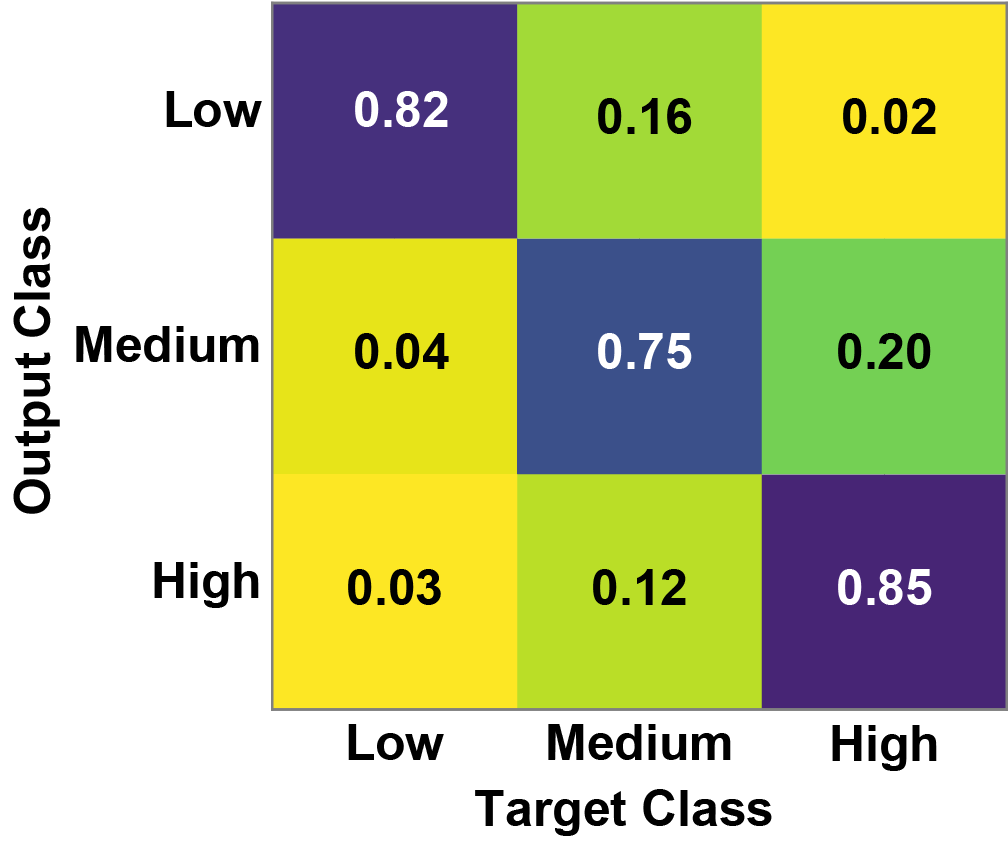}
\caption{Confusion Matrix for Three-Level Trust Classification Using All Physiological Signals with CP Guidance.}
\label{fig_5}
\end{figure}

Fig. \ref{fig_5} presents the confusion matrix for three-level classification when fusing all features with CP. In this matrix, columns represent ground truth classes, while rows represent output classes. Diagonal elements indicate the percentage of correctly classified samples for each class. Our results demonstrate successful differentiation of human trust across three levels, validating the capability of our PPTP model. The medium category showed lower accuracy than the other two categories, which we will discuss in detail in Section IV-E.

\subsection{Effectiveness of CP Guidance Across Signal Combinations}
\begin{figure}[tb]
\centering
\includegraphics[width=2.5in]{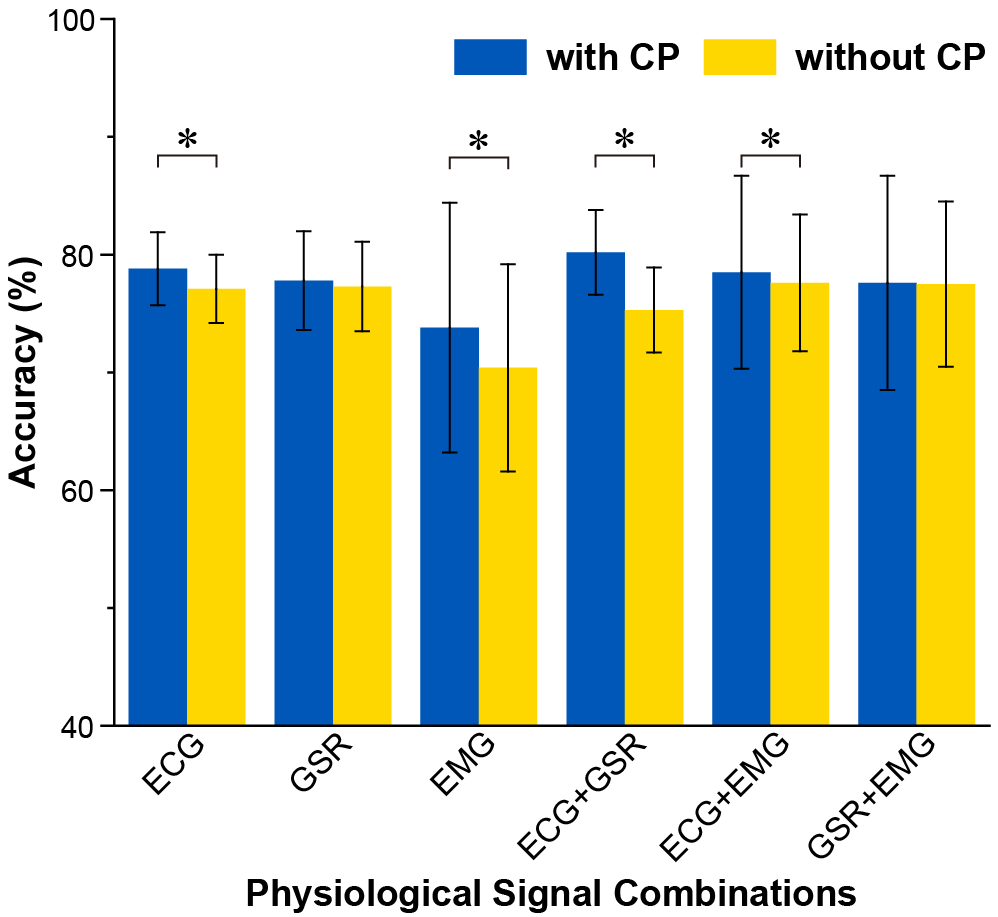}
\caption{Classification Accuracy With and Without CP Guidance Across Different Signal Combinations.}
\label{fig_6}
\end{figure}

Based on the higher accuracy exhibited when using CP guidance with all modality physiological signals, we calculated the accuracy of different physiological signal combinations with and without CP evaluation to validate CP's effectiveness. Fig. \ref{fig_6} shows the average classification accuracy and variability across all participants. Results demonstrate that introducing CP information significantly improves the performance of several physiological signal combinations, validating the effectiveness of our designed CP metric for trust prediction. As shown in TABLE \ref{tab:tab2}, the model achieves more than 77\% accuracy when any one physiological signal modality is missing, particularly 80.2\% accuracy when EMG is absent. The model maintains more than 73\% accuracy even when two physiological signals are missing. 

\begin{table}[h]
\caption{Comparison of Accuracy and F1 Score for Various Physiological Signal and CP Combinations.}
\label{tab:tab2}
\begin{center}
\begin{tabular}{|l|cc|cc|}
\hline
\multirow{2}{*}{\textbf{Signals}} & \multicolumn{2}{c|}{\textbf{Accuracy}} & \multicolumn{2}{c|}{\textbf{F1 score}} \\
\cline{2-5}
& Mean & Std & Mean & Std \\
\hline
CP + ECG & 78.8 & 3.1 & 73.7 & 6.0 \\
CP + GSR & 77.8 & 4.2 & 71.7 & 5.2 \\
CP + EMG & 73.8 & 10.6 & 63.6 & 12.2 \\
CP + ECG + GSR & 80.2 & 3.6 & 75.5 & 7.6 \\
CP + ECG + EMG & 78.5 & 8.2 & 73.7 & 10.5 \\
CP + EMG + GSR & 77.6 & 9.1 & 71.9 & 11.4 \\
ECG + GSR & 75.3 & 3.6 & 68.4 & 7.0 \\
ECG + EMG & 77.6 & 5.8 & 71.3 & 8.1 \\
EMG + GSR & 77.5 & 7.0 & 59.2 & 7.5 \\
\hline
\end{tabular}
\end{center}    
\end{table}

These results demonstrate the model's robustness when signal modalities are unavailable. Combined with the unimodal signal classification results, these findings further validate the effectiveness of ECG and GSR signals for dynamic trust prediction, as well as the enhancement provided by CP information when fused with physiological signals.

\subsection{Comparison with State-of-the-art Methods}
TABLE \ref{tab:tab3} compares our method's classification results with several multimodal sequential prediction baselines using all physiological signals and CP. Our method achieves the best performance among the baselines, exceeding Transformer by 6.7\% and being the only method to achieve over 80\% accuracy.

Among the compared methods, CPM-Nets \cite{zhang2020deep} is a multi-view classification approach that has demonstrated effectiveness in cognitive load prediction based on physiological signals. To address scale differences between physiological signals and CP for baseline methods, we implemented an embedding module that maps CP to the same dimensional space as physiological signals before concatenating them as a unified input.

\begin{table}[h]
\caption{Performance Comparison Between PPTP and State-of-the-art Baseline Methods.}
\label{tab:tab3}
\begin{center}
\begin{tabular}{|c|c|c|c|c|c|}
\hline
\multirow{2}{*}{\textbf{Accuracy}} & \multicolumn{5}{c|}{\textbf{Method}} \\ \cline{2-6}
& CPM-Nets & RNN & LSTM & Transformer & Ours \\ \hline
Mean & 66.2 & 63.7 & 71.6 & 74.4 & \textbf{81.1} \\ \hline
Std & 6.9 & 3.3 & 6.4 & 4.5 & 3.6 \\ \hline

\end{tabular}
\end{center}
\end{table}

\subsection{Seven-Level vs. Three-Level Trust Classification Performance}
To further validate our method's effectiveness and the correlation between trust levels and physiological signals, we conducted high-resolution trust classification experiments. We replaced the three-level trust labels (low, medium, high) with seven-level Muir scale scores, with Labels 1 through 7 representing trust levels from "very low" to "very high."

\begin{table}[tb]
\caption{Comparison for Three-Level vs. Seven-Level Trust Classification Tasks.}
\begin{center}
\begin{tabular}{|l|cc|cc|}
\hline
\multirow{2}{*}{\textbf{Signals}} & \multicolumn{2}{c|}{\textbf{Accuracy}} & \multicolumn{2}{c|}{\textbf{F1 score}} \\
\cline{2-5}
& Mean & Std & Mean & Std \\
\hline
All physiological signals + CP & 81.1 &3.6 & 75.9 &7.1\\
All physiological signals & 78.5 &3.4 & 73.1 &7.4\\
All physiological signals + CP (7-level)& 74.3 &6.5 & 69.4 &7.8\\
All physiological signals (7-level)& 72.9 &6.9 & 67.4 &8.4\\
\hline
\end{tabular}
\end{center}
\label{tab:tab4}
\end{table}

\begin{figure}[tb]
\centering
\includegraphics[width=2in]{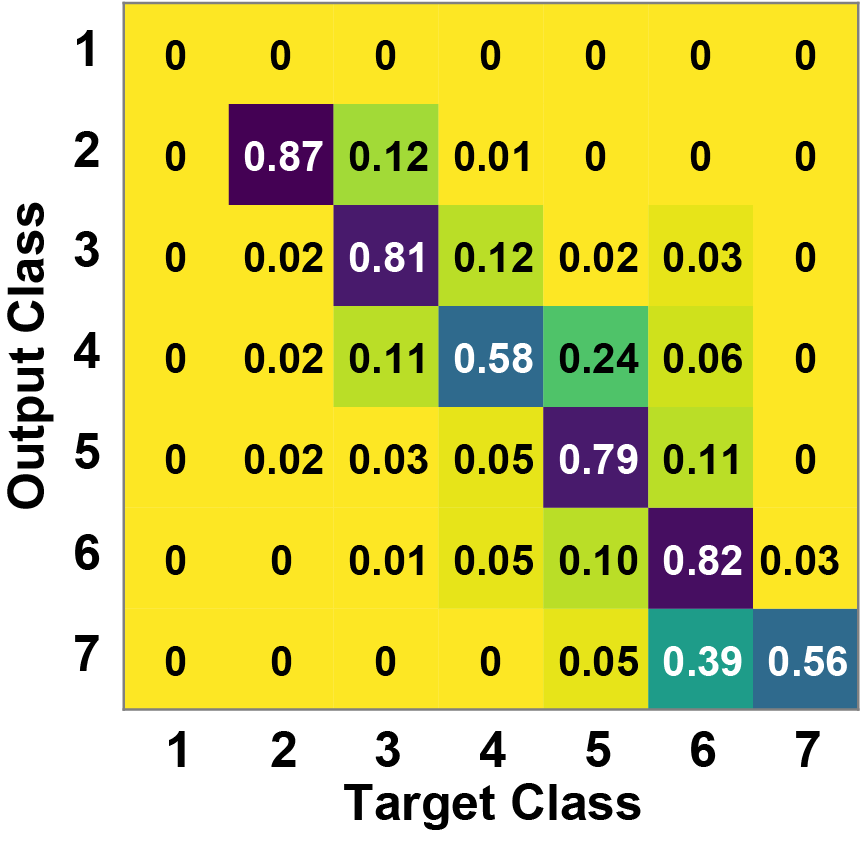}
\caption{Confusion Matrix for High-Resolution Seven-Level Trust Classification with CP Guidance.}
\label{fig_7}
\end{figure}

As shown in TABLE \ref{tab:tab4}, CP guidance significantly improved multimodal physiological signal classification performance in seven-level classification, achieving 74.3\% accuracy compared to 72.9\% without CP. Fig. \ref{fig_7} presents the average confusion matrix for seven-level classification when fusing all physiological signals with CP. Evidently, except for the medium class (Label 4) and extreme class (Label 7), all other classes exhibit satisfactory classification accuracy. Misclassifications for these two classes mostly occur in neighboring classes, verifying our model's robustness across multi-resolution ratio modes in the experimental design.

Two possible explanations exist for the lower accuracy in medium level classification. First, participants' physiological signal features may be less pronounced in neutral stance situations (label 4) compared to more extreme situations, resulting in less significant signal features. This aligns with the observation that classification accuracy increases as trust levels deviate from the medium level (except in extreme situations), a pattern also observed in three-level classification. Second, the neutral boundary is inherently ambiguous, with label 4 easily confused with labels 3 and 5 in subjective evaluations, increasing categorization difficulty. This is consistent with the observation that false positives for label 4 are mostly concentrated in label 3 (0.11) and label 5 (0.24).

Regarding the lower classification accuracy for extreme labels, two potential explanations can be offered. First, physiological signals may reach saturation when a person experiences extremely high trust, making differentiation from label 6 difficult. This aligns with the observation that false positives for label 7 are mostly concentrated in label 6 (0.39). Second, individual differences in both subjective reporting and objective physiological responses when expressing "very high trust" may result in diverse physiological signal patterns. As the highest trust assessment level, the subjective assessment boundary is more ambiguous and easily confused with label 6.

\section{CONCLUSION}
In this study, we investigated human trust prediction by integrating physiological signals with collaboration performance assessment in HRC construction tasks. Our approach incorporated ECG, GSR, and EMG signals as physiological inputs, utilizing collaboration performance as guiding information in a novel PPTP framework. This architecture effectively fused and aligned multimodal signals to accurately classify trust levels. Results demonstrated the PPTP model's outstanding performance, achieving over 80\% accuracy in three-level trust classification. We also accomplished high-resolution seven-level trust classification for the first time with 74.3\% accuracy. Our comprehensive analysis of individual signals and various combinations confirmed the reliability of physiological signals when augmented with CP assessment, validating both the individual components and overall methodology. This research contributes new insights to trust prediction in human-robot collaboration and establishes a foundation for cross-modal information fusion that enhances prediction validity. The approach offers promising applications for real-time trust monitoring in collaborative robotics, particularly in industry settings where humans and robots function as equal teammates.


\bibliographystyle{ieeetr}
\bibliography{references}

\end{document}